\documentclass[fleqn,usenatbib]{mnras}
\hypersetup{draft} 

\usepackage{newtxtext,newtxmath}

\usepackage[T1]{fontenc}
\usepackage{ae,aecompl} 
\usepackage{lscape}

\usepackage{graphicx}	
\usepackage{amsmath}	
\usepackage{amssymb}	
\usepackage[export]{adjustbox}
\usepackage{subfig}

\newcommand{\kepler}{{\it Kepler}}
\newcommand{\corot}{{\it CoRoT}}

\newcommand{\NGTS}{NGTS}
\newcommand{\TESS}{{\it TESS}}
\newcommand{\gpe}{{GP-EBOP}}
\newcommand{\LSO}{La Silla Observatory}

\newcommand{\Euler}{Eulercam}
\newcommand{\LCO}{LCO\,1m}
\newcommand{\SAAO}{SAAO}
\newcommand{\SHOC}{SHOC}
\newcommand{\SHOCa}{SHOCa}

\newcommand{\kms}{km\,s$^{-1}$}
\newcommand{\ms}{m\,s$^{-1}$}
\newcommand{\masy}{mas\,yr$^{-1}$}
\newcommand{\mpl}{\mbox{M$_{p}$}}
\newcommand{\rpl}{\mbox{R$_{p}$}}
\newcommand{\mstar}{\mbox{M$_{*}$}}
\newcommand{\rstar}{\mbox{R$_{*}$}}
\newcommand{\mjup}{\mbox{M$_{J}$}}

\newcommand{\msun}{\mbox{M$_{\odot}$}}
\newcommand{\rsun}{\mbox{R$_{\odot}$}}
\newcommand{\rearth}{R$_{\oplus}$}
\newcommand{\mearth}{M$_{\oplus}$}
\newcommand{\gccc}{g\,cm$^{-3}$}

\newcommand{\teff}{$T_{\rm eff}$}
\newcommand{\teq}{$T_{\rm eq}$}
\newcommand{\logg}{$\log g$}
\newcommand{\FeH}{[Fe/H]}
\newcommand{\vmac}{$v_{\rm mac}$}

 \newcommand{\Nstar}{NGTS-4}
 \newcommand{\Nkamp}{\mbox{$13.7 \pm 1.9$}}
 \newcommand{\Ngamma}{\mbox{$111.2 \pm 0.2 $}}
 \newcommand{\NRA}{\mbox{$05^{\rmn{h}} 58^{\rmn{m}} 23\fs76$}} 
 \newcommand{\NDec}{\mbox{$-30\degr 48\arcmin 42\farcs 49$}} 
 \newcommand{\Ntwomass}{\mbox{$05582375$-$3048424$}} 

 \newcommand{\NpropRA}{\mbox{$-16.881\pm0.034$}} 
 \newcommand{\NpropDec}{\mbox{$-7.371\pm0.036$}} 
 \newcommand{\Ngaiagamma}{\mbox{$110.5\pm5.5$}} 
 \newcommand{\Nparallax}{\mbox{$3.536\pm0.023$}}
 \newcommand{\Ngaia}{\mbox{$2891248292906892032$}}
 \newcommand{\Ndist}{\mbox{$282.6\pm1.8$}}
 
\newcommand{\NstarradiusSPECIES}{\mbox{$0.84\pm0.01$}}
\newcommand{\NstarmassSPECIES}{\mbox{$0.75\pm0.02$}}
\newcommand{\NteffSPECIES}{\mbox{$5143\pm100$}}
\newcommand{\NloggSPECIES}{\mbox{$4.5\pm0.1$}}
\newcommand{\NmetalSPECIES}{\mbox{$-0.28\pm0.10$}}
\newcommand{\NvsiniSPECIES}{\mbox{$2.75\pm0.58$}}
\newcommand{\NvmacSPECIES}{\mbox{$1.40\pm0.58$}}

 \newcommand{\Nstardensity}{\mbox{$1.79\pm0.14$}}

 \newcommand{\NVmag}{$13.14 \pm 0.03$}
 \newcommand{\NBmag}{$13.95 \pm 0.05$}
  \newcommand{\Ngmag}{$13.48 \pm 0.02$}
 \newcommand{\Nrmag}{$12.91 \pm 0.07$}
 \newcommand{\Nimag}{$12.64 \pm 0.09$}
 \newcommand{\NGAIAmag}{$12.91 $}
 \newcommand{\NGAIABPmag}{$13.36 $}
 \newcommand{\NGAIARPmag}{$12.31 $}
 \newcommand{\NNmag}{$12.59 \pm 0.01$}
 \newcommand{\NJmag}{$11.58 \pm 0.02$}
 \newcommand{\NHmag}{$11.14 \pm 0.02$}
 \newcommand{\NKmag}{$11.07 \pm 0.02$}
 \newcommand{\NWmag}{$11.03 \pm 0.02$}
 \newcommand{\NWWmag}{$11.09 \pm 0.02$}
 \newcommand{\NWWWmag}{$10.98 \pm 0.11$}
 \newcommand{\Nplanet}{NGTS-4b}
 \newcommand{\Nperiod}{\mbox{$1.3373508 \pm 0.000008$}}
 \newcommand{\Nperiodshort}{\mbox{$1.33734$}}
 \newcommand{\Nduration}{\mbox{$1.80\pm0.10$}}%
 \newcommand{\Ntc}{\mbox{$2457607.9975 \pm 0.0034$}}
 \newcommand{\Nmass}{\mbox{$20.6\pm3.0$}}%
 \newcommand{\Nradius}{\mbox{$3.18\pm0.26$}}%
 \newcommand{\Ndensity}{\mbox{$3.45\pm0.95$}}%

 \newcommand{\NTeq}{\mbox{$1650\pm 400$}}%

 \newcommand{\Nau}{\mbox{$0.019\pm0.005$}}%
 \newcommand{\Naoverr}{\mbox{$4.79\pm1.21$}}%

\newcommand{\Ndepth}{\mbox{$0.13\pm0.02$}}%
\newcommand{\Ninclination}{\mbox{$82.5\pm5.8$}}

\newcommand{\Nbaselinenights}{272}
\newcommand{\Nbaselinestart}{2016 August 06}
\newcommand{\Nbaselineend}{2017 May 05}
\newcommand{\Nnimages}{190\,696}
\newcommand{\Nguidingrmspixels}{0.051}
\newcommand{\progid}{0101.C-0623(A)}
\newcommand{\noRV}{14}
\newcommand{\mask}{K0}

\title[\Nplanet]{\Nplanet: A sub-Neptune Transiting in the Desert}

\author[R.~G.~West et al.]{
\parbox{\textwidth}{
Richard~G.~West,$^{1, 2}$\thanks{E-mail: \href{richard.west@warwick.ac.uk}{richard.west@warwick.ac.uk}}
Edward~Gillen,$^{3,\dagger}$
Daniel~Bayliss,$^{1, 2}$
Matthew R. Burleigh,$^{5}$
Laetitia Delrez,$^{3}$
Maximilian~N.~G\"unther,$^{3}$
Simon~T.~Hodgkin,$^{7}$
James~A.~G.~Jackman,$^{1, 2}$
James~S.~Jenkins,$^{8, 9}$
George King,$^{1, 2}$
James McCormac,$^{1, 2}$
Louise~D.~Nielsen,$^{4}$
Liam Raynard,$^{5}$
Alexis~M.~S.~Smith,$^{6}$
Maritza~Soto,$^{8}$
Oliver Turner,$^{4}$ 
Peter~J.~Wheatley,$^{1,2}$
Yaseen~Almleaky,$^{15,16}$
David J. Armstrong,$^{1, 2}$
Claudia Belardi,$^{5}$
Fran\c{c}ois Bouchy,$^{4}$
Joshua T. Briegal,$^{3}$
Artem~Burdanov,$^{14}$
Juan Cabrera,$^{6}$
Sarah L. Casewell,$^{5}$
Alexander Chaushev,$^{5, 1, 2}$
Bruno Chazelas,$^{4}$
Paul Chote,$^{1, 2}$
Benjamin~F.~Cooke,$^{1, 2}$
Szilard Csizmadia,$^{6}$
Elsa~Ducrot,$^{14}$
Philipp~Eigm\"uller,$^{6,10}$
Anders Erikson,$^{6}$
Emma~Foxell,$^{1,2}$
Boris~T.~G\"ansicke,$^{1,2}$
Micha\"el Gillon,$^{14}$
Michael R.~Goad,$^{5}$
Emmanu\"el~Jehin,$^{14}$
Gregory Lambert,$^{3}$
Emma~S.~Longstaff,$^{5}$ 
Tom Louden,$^{1, 2}$
Maximiliano~Moyano,$^{13}$
Catriona~Murray,$^{3}$
Don~Pollacco,$^{1, 2}$
Didier~Queloz,$^{3}$
Heike~Rauer,$^{6, 10, 11}$
Sandrine~Sohy,$^{14}$
Samantha~J.~Thompson,$^{3}$
St\'{e}phane~Udry,$^{4}$
Simon.~R.~Walker,$^{1, 2}$
Christopher~A.~Watson$^{12}$
}
\\
\\
$^{1}$Centre for Exoplanets and Habitability, University of Warwick, Gibbet Hill Road, Coventry CV4 7AL, UK\\
$^{2}$Dept.\ of Physics, University of Warwick, Gibbet Hill Road, Coventry CV4 7AL, UK\\
$^{3}$Astrophysics Group, Cavendish Laboratory, J.J. Thomson Avenue, Cambridge CB3 0HE, UK\\
$^{4}$Observatoire de Gen{\`e}ve, Universit{\'e} de Gen{\`e}ve, 51 Ch. des Maillettes, 1290 Sauverny, Switzerland\\
$^{5}$Department of Physics and Astronomy, Leicester Institute of Space and Earth Observation, University of Leicester, LE1 7RH, UK\\
$^{6}$Institute of Planetary Research, German Aerospace Center, Rutherfordstrasse 2, 12489 Berlin, Germany\\
$^{7}$Institute of Astronomy, Cambridge University, Madingley Road, Cambridge CB3 0HA, UK\\
$^{8}$Departamento de Astronomia, Universidad de Chile, Casilla 36-D, Santiago, Chile\\
$^{9}$ Centro de Astrof\'isica y Tecnolog\'ias Afines (CATA), Casilla 36-D, Santiago, Chile.\\
$^{10}$Center for Astronomy and Astrophysics, TU Berlin, Hardenbergstr. 36, D-10623 Berlin, Germany\\
$^{11}$Institute of Geological Sciences, FU Berlin, Malteserstr. 74-100, D-12249 Berlin, Germany\\
$^{12}$Astrophysics Research Centre, School of Mathematics and Physics, Queen's University Belfast, BT7 1NN Belfast, UK\\
$^{13}$Instituto de Astronom\'ia, Universidad Cat\'olica del Norte, Angamos 06010, 1270709, Antofagasta, Chile\\
$^{14}$Space sciences, Technologies and Astrophysics Research (STAR) Institute, Universit\'e de Li\`ege, 4000 Li\`ege, Belgium\\
$^{15}$Space and Astronomy Department, Faculty of Science, King Abdulaziz University, 21589 Jeddah, Saudi Arabia\\
$^{16}$King Abdullah Centre for Crescent Observations and Astronomy, Makkah Clock, Mecca 24231, Saudi Arabia\\
$^{\dagger}$\,Winton Fellow
}
\date{Submitted 2018 August 30}

\pubyear{2018}

\begin{document}
\label{firstpage}
\pagerange{\pageref{firstpage}--\pageref{lastpage}}
\maketitle

\begin{abstract}
We report the discovery of \Nplanet, a sub-Neptune-sized  planet transiting a 13th magnitude K-dwarf in a 1.34\,d orbit. \Nplanet\ has a mass M=\Nmass\,\mearth\ and radius R=\Nradius\,\rearth, which places it well within the so-called ``Neptunian Desert''. The mean density of the planet (\Ndensity\,\gccc) is consistent with a composition of 100\,\% H$_2$O or a rocky core with a volatile envelope. NGTS-4b is likely to suffer significant mass loss due to relatively strong EUV/X-ray irradiation. Its survival in the Neptunian desert may be due to an unusually high core mass, or it may have avoided the most intense X-ray irradiation by migrating after the initial activity of its host star had subsided. With a transit depth of \Ndepth\%, \Nplanet\ represents the shallowest transiting system ever discovered from the ground, and is the smallest planet discovered in a wide-field ground-based photometric survey.
\end{abstract}

\begin{keywords}
techniques: photometric, stars: individual: \Nstar, planetary systems, planets and satellites: detection
\end{keywords}






\section{Introduction}
\label{sec:intro}
Exoplanet population statistics from the Kepler mission reveals a scarcity of short period Neptune-sized planets \citep{szabo2011, mazeh2016, Fulton2018}.  This so-called ``Neptunian Desert" is broadly defined as the lack of exoplanets with masses around 0.1\,\mjup\  and periods less than 2--4 days \citep{szabo2011}. As Neptune-sized planets should be easier to find in short period orbits, and many Neptunes have been discovered with longer orbits from surveys such as \corot\ and \kepler , this does not appear to be an observational bias.  Ground-based surveys, which have uncovered the bulk of known hot Jupiters, have not uncovered these short-period Neptunes.  However this may be due to the fact that such exoplanets produce transits too shallow for most ground-based surveys to detect.

The physical mechanisms that result in the observed Neptunian Desert are currently unknown, but have been suggested to be due to a different formation mechanism for short period super-Earth, and Jovian exoplanets, similar to the reasons for the brown dwarf desert (e.g. \citealt{grether2006}). Alternatively, the dearth may be due to a mechanism stopping planetary migration. This may be a sudden loss of density within the accretion disk, or mass removed from the exoplanet via Roche lobe overflow \citep{kurokawa2014} or stellar X-ray/EUV insolation \citep{lopez2014} and evaporation of the atmosphere \citep{lecav2007}.

\citet{Owen2018} investigated causes of the high mass/large radius and low mass/small radius boundaries of the desert. They showed that while X-ray/EUV photoevaporation of sub-Neptunes can explain the low mass/small radius boundary, the high mass/large radius boundary better corresponds to the tidal disruption barrier for gas giants undergoing high eccentricity migration. Their findings were consistent with the observed triangular shape of the desert, since photoevaporation is more prolific at shorter orbital periods, likewise more massive gas giants can tidally circularise closer to their stellar hosts.

Due to their shallow transits, Neptune-sized planets ($\approx$4\,\rearth) have largely eluded wide-field ground-based transit surveys such as WASP \citep{waspproject}, HATNet \citep{hatproject}, HATSouth \citep{hatsproject}, and KELT \citep{keltproject1, keltproject2}. The notable exception is HAT-P-11b \citep{hat11}, which has a radius of just $4.71\pm0.07$\,\rearth. One other system worthy of note is the multi-planet system TRAPPIST-1 \citep{gillon2016}, of which three of the Earth-sized planets were discovered from ground, however they orbit a late M-dwarf and their transit depths are in the range 0.6--0.8\% (5--6 times larger than the depth of NGTS-4b), and surveys such as TRAPPIST and MEarth \citep{Nutzman2008,Irwin2009} have specifically targeted M-dwarfs in order to maximise the detectability of small planets.

We present the discovery of a new sub-Neptune-sized (R=\Nradius\,\rearth) planet transiting a K-dwarf (\mbox{$m_v=13.1$}\,mag) in a P=\Nperiodshort\,d orbit from the Next Generation Transit Survey (\NGTS) survey. In Sect.~\ref{sec:ngts} we describe the NGTS discovery data.  In Sect.~\ref{sec:phfu} we describe our campaign of photometric follow-up on 1\,m-class telescopes.  In Sect.~\ref{sec:spec} we detail our spectroscopic follow-up including the mass determination via radial velocity monitoring.  In Sect.~\ref{sec:analysis} we discuss our analysis of the stellar parameters and describe the global modelling process. In Sect.~\ref{sec:discussion} we discuss the discovery in context with other planets in this mass/radius/period regime.  Finally we finish with our conclusions in Sect.~\ref{sec:conclusions}.


\section{Discovery Photometry From \NGTS}
\label{sec:ngts}

\Nstar~was observed using a single NGTS camera over a \Nbaselinenights~night baseline between \Nbaselinestart~and \Nbaselineend . The survey has operated at ESO's Paranal observatory since early 2016 and consists of an array of twelve roboticized 20\,cm telescopes. The facility is optimised for detecting small planets around K and early M stars \citep{Wheatley2018,McCormac2017,2013EPJWC..4713002W,Chazelas2012}.

A total of \Nnimages~images were obtained, each with an exposure time of $10$ s. The data were taken using the custom \NGTS~filter (550--927\,nm) and the telescope was auto-guided using an improved version of the DONUTS auto-guiding algorithm \citep{McCormac2013}. The RMS of the field tracking errors was \Nguidingrmspixels~pixels ($0.26''$) over the \Nbaselinenights~night baseline. The data were reduced and aperture photometry was extracted using the CASUTools\footnote{\url{http://casu.ast.cam.ac.uk/surveys-projects/software-release}} photometry package. A total of 185\,840 valid data-points were extracted from the raw images. The data were then de-trended for nightly trends, such as atmospheric extinction, using our implementation of the SysRem algorithm \citep{tamuz05, colliercameron2006}. We refer the reader to \citet{Wheatley2018} for more details on the NGTS facility and the data acquisition and reduction processes. 

\begin{figure}
	\includegraphics[width=8.5cm]{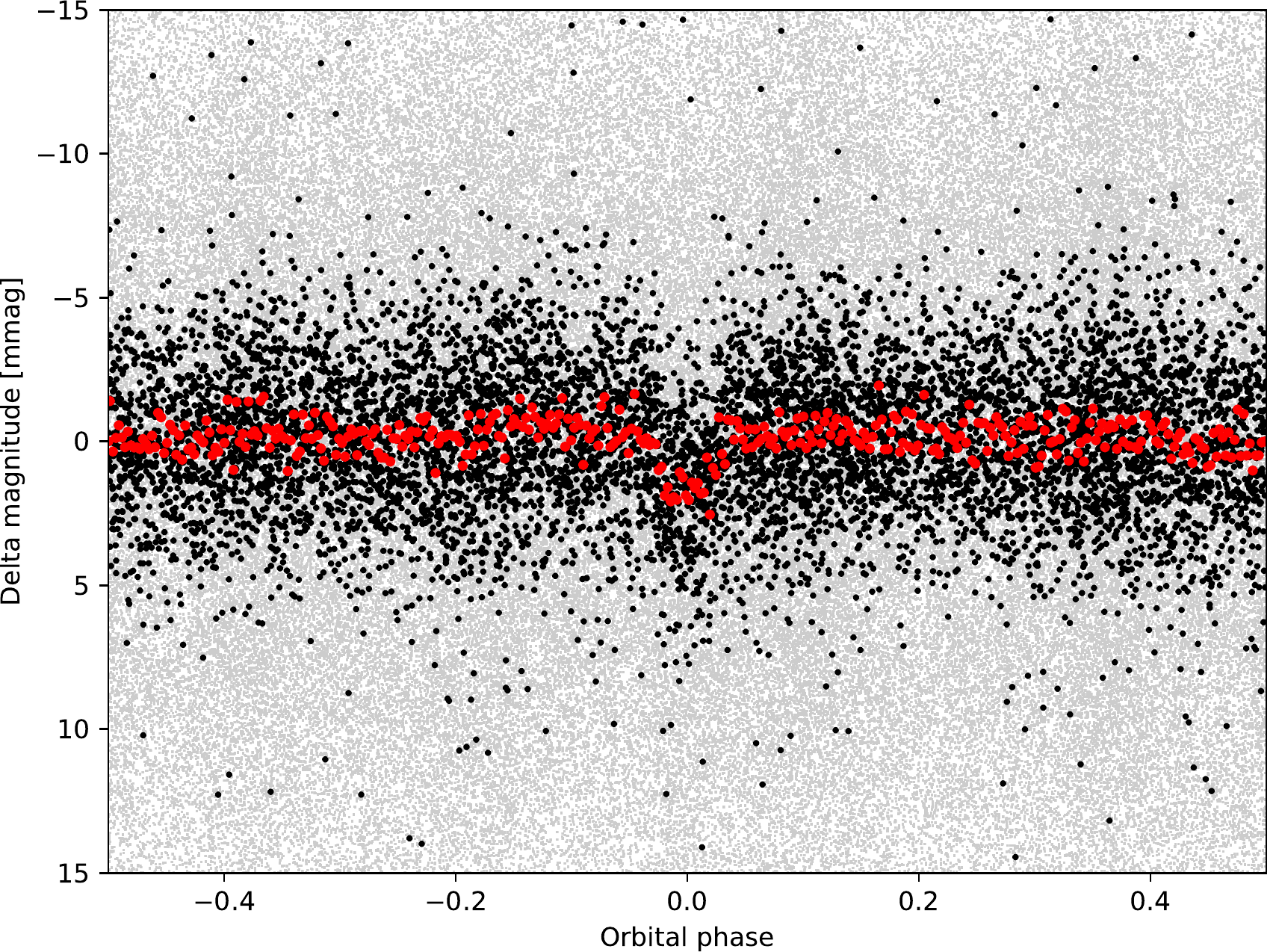}
    \caption{The \NGTS\ discovery light curve of \Nplanet. The data are shown phase-folded on the orbital period \Nperiodshort\,d. The grey points show the unbinned 10\,s cadence data, the black dots are these data binned in linear time to a cadence of 5\,min then phase-folded, and the red points are the unbinned data phase-folded then binned in phase to an equivalent cadence of 5\,min.} 
    \label{fig:ngtsphot}
\end{figure}

The complete dataset was searched for transit-like signals using {\sc orion}, an optimized implementation of the box-fitting least-squares (BLS) algorithm \citep{kovacs2002, colliercameron2006}, and a $\sim$0.2\,\% signal was detected at a period of \Nperiodshort\,d. The \NGTS{} photometry, phase-folded to this period, is shown in Figure~\ref{fig:ngtsphot}. A total of 28 transits are covered fully or partially by the NGTS dataset.

We find no evidence for a secondary eclipse or out-of-transit variation, both of which would indicate an eclipsing binary system.  We used the centroid vetting procedure detailed in \citet{Guenther17b} to check for contamination from background objects, and verify that the transit seen was not a false positive. This test is able to detect shifts in the photometric centre-of-flux during transit events at the sub-milli-pixel level. It can identify blended eclipsing binaries at separations below 1\arcsec, well below the size of individual \NGTS\ pixels (5\arcsec). We find no signs of a centroiding variation during the transit events of \Nstar. 

Based on the \NGTS\ detection and the above vetting tests, \Nstar\ was followed-up with further photometry and spectroscopy to confirm the planetary nature of the system and measure the planetary parameters. A sample of the full discovery photometry and follow-up data is given in Table~\ref{tab:allphot}, the full data are available in machine-readable format from the online journal. 

\begin{table}
	\centering
	\caption{Summary of photometry of \Nstar. This table contains all the photometry of \Nstar{} from \NGTS, \Euler, \LCO, and \SHOCa. A portion is shown here for guidance. The full table is available in a machine-readable format from the online journal.}
	\label{tab:allphot}
	\begin{tabular}{ccccc}
	Time            & Relative & Flux  & Filter & Instrument\\
    (HJD-2450000)   & flux     & error &        &           \\
	\hline
    7287.76965278 & 1.00973 & 0.01865 & NGTS & NGTS \\
    7287.76979167 & 1.01170 & 0.01861 & NGTS & NGTS \\
    7287.76994213 & 0.99010 & 0.01855 & NGTS & NGTS \\
    7287.77008102 & 1.00987 & 0.01860 & NGTS & NGTS \\
    7287.77023148 & 0.99255 & 0.01862 & NGTS & NGTS \\
    7287.77037037 & 0.99335 & 0.01858 & NGTS & NGTS \\
    7287.77052083 & 0.99226 & 0.01856 & NGTS & NGTS \\
    7287.77068287 & 0.99369 & 0.01859 & NGTS & NGTS \\
    7287.77082176 & 0.98936 & 0.01857 & NGTS & NGTS \\
    7287.77097222 & 1.02166 & 0.01867 & NGTS & NGTS \\
    ...           & ...     & ...     & ...  & ...  \\
	\hline
	\end{tabular}
\end{table}


\section{Photometric follow-up}
\label{sec:phfu}
Confirming such a shallow transit signal from the ground is challenging, even given some of the best 1\,m-class telescopes available for precision time-series photometry.  We therefore undertook a campaign of photometric follow-up using four different facilities as set out in this section. A summary of the photometric follow-up observations is given in Table~\ref{tab:followup_phot}, and the full data are available in machine-readable format in the online journal.  The de-trended data are plotted in Figure~\ref{fig:global_fit_fup_phot} (see Sect.~\ref{sub:global} for a description of the de-trending that has been applied to the data in these plots).

\subsection{SHOC photometry}
\label{sub:saaophot}

Our first follow-up photometry of \Nstar\ was carried out at the South African Astronomical Observatory (\SAAO) on 2017 November 27, with the 1.0\,m telescope and one of the three frame-transfer CCD Sutherland High-speed Optical Cameras \citep[SHOC]{Coppejans:2013gx}, specifically SHOC'n'awe. The \SHOC~cameras on the 1\,m telescope have a pixel scale of $0.167$\arcsec/pixel, which is unnecessarily fine for our observations, hence we binned the camera $4\times4$ pixels in the X and Y directions. All observations were obtained in focus, using a $V$ filter and an exposure time of 30\,s. The field of view of the SHOC instruments on the 1\,m is $2.85'\times 2.85$\arcmin, which allowed for one comparison star of similar brightness to the target to be simultaneously observed. 

The data were bias and flat-field corrected via the standard procedure using the CCDPROC package \citep{2015ascl.soft10007C} in python. Aperture photometry was extracted for \Nstar{} and the comparison star using the SEP package \citep{Barbary16, 1996A&AS..117..393B} and the sky background was measured and subtracted using the SEP background map. We also performed aperture photometry using the Starlink package {\sc autophotom}. We used a 4 pixel radius aperture that maximised the signal/noise, and the background was measured in an annulus surrounding this aperture. The comparison star was then used to perform differential photometry on the target.  Both photometry methods successfully detected a complete transit of \Nplanet\, despite the observations being partially effected by thin cirrus during the transit. 

\Nstar\ was observed again at \SAAO\ with the 1\,m telescope and the SHOC'n'awe instrument at the end of astronomical twilight on 2018 April 22. On this occasion sky conditions were excellent, with sub-arcsecond seeing throughout and a minimum of $\approx0.6$\arcsec\ recorded. On this occasion the observations were made using an $I$ filter. Initially, an exposure time of 5\,s was used, but after the first 30\,minutes this was reduced to 2\,s as the target's flux was uncomfortably close to the non-linear regime of the CCD. These data were also reduced and analysed as described above, and the transit egress was clearly detected (Figure~\ref{fig:global_fit_fup_phot}).

\subsection{LCO 1\,m}
\label{sub:lco}
We monitored transit events of \Nplanet\ using the Las Cumbres Observatory (LCO) 1\,m global telescope network \citep{brown2013}.  All observations were taken using the Sinistro cameras, which give an a $26.5'\times26.5'$ field of view with a plate-scale of 0.389\arcsec/pixel.  Exposure times were set to 180\,s, with a defocus of 2\,mm in order to ensure we did not saturate \Nstar\ and light was spread over a larger number of detector pixels.  We used the $i$-band filter and the standard $1 \times 1$ binning readout mode.  In total six events were monitored with the LCO 1\,m telescopes from the sites in Chile and Australia.  A full list of these events along with details of each observation are set out in Table~\ref{tab:followup_phot}.

Raw images were reduced to calibrated frames using the standard LCO ``Banzai" pipeline. Aperture photometry was extracted for \Nstar{} and the 7 comparison stars using the {\sc sep} package \citep{Barbary16, 1996A&AS..117..393B} and the sky background was measured and subtracted using the {\sc sep} background map. The resulting light-curve shows the signature of a full transit (Figure~\ref{fig:global_fit_fup_phot}).

\subsection{Speculoos}
\label{sub:speculoos}

We monitored a transit event using the SPECULOOS-South facility (\citealt{burdanov2017,Delrez18}) at Paranal Observatory in Chile on the night of 2018 April 15, taking advantage of the telescope commissioning period. SPECULOOS-South consists of four robotic 1-meter Ritchey-Chretien telescopes, and we were able to utilize two of these (Europa and Callisto) to observe the transit event.  Given the shallowness of the targeted transit, we opted to maximise the flux from the early K host star and chose an I+z filter for both telescopes. SPECULOOS-South is equipped with a deep-depletion $2{\rm k}\times2{\rm k}$ CCD camera with a field-of-view of $12'\times12'$ (0.35\arcsec/pixel).

The images were calibrated using standard procedures (bias, dark, and flat-field correction) and photometry was extracted using the \textsc{IRAF/DAOPHOT} aperture photometry software \citep{stetson1987}, as described by \citet{gillon2013}. For each observation, a careful selection of both the photometric aperture size and stable comparison stars was performed manually to obtain the most accurate differential light curve of \Nstar{}. The signature of a full transit is evident in the light-curves from both telescopes (Figure~\ref{fig:global_fit_fup_phot}).

\subsection{Eulercam}
\label{sub:eulercam}

Two transits of \Nstar{} were observed with \Euler~on the 1.2\,m Euler Telescope at \LSO. The observations took place on the nights beginning 2018 April 15 and 2018 April 19. Both transits were observed using the same broad NGTS filter that was used to obtain the discovery photometry. For the first observation a total of 193 images were obtained using a 40\,s exposure and 0.1\,mm defocus.  For the second observation a total of 140 images were obtained using a 55\,s exposure time and 0.1\,mm defocus.  

The data were reduced using the standard procedure of bias subtraction and flat-field correction. Aperture photometry was performed with the PyRAF implementation of the phot routine.  PyRAF was also used to extract information useful for de-trending; X- and Y-position, FWHM, airmass and sky background of the target star. The comparison stars and the photometric aperture radius were chosen in order to minimise the RMS in the scatter out of transit. Additional checks were made with different comparison star ensembles, aperture radii and with stars in the FOV expected to show no variation. This was to ensure the transit signal was not an artefact of these choices. The resulting light-curves are plotted in Figure~\ref{fig:global_fit_fup_phot}, showing a detection of a full transit signature in the data from 2018 April 19, though the detection in the data from 2018 April 15 is marginal at best.


\begin{table*}
	\centering
	\caption{A summary of the follow-up photometry of \Nstar{}}
	\label{tab:followup_phot}
	\begin{tabular}{ccccccccc} 
		Night & Site & Instrument & N$_{\mathrm{images}}$ & Exptime & Binning & Filter & Comment \\
        &  & & & (seconds) & & & \\
		\hline
2017 Nov 27 & SAAO & \SHOC   & 360   & 30 & 4$\times$4 & V & full transit    \\
2018 Apr 15 & CTIO & \LCO     & 31  & 180 & 1$\times$1 & i & mid+egress \\
2018 Apr 15 & Paranal & Speculoos/Callisto & 471   & 12 & 1$\times$1 & I+z & full transit  \\
2018 Apr 15 & Paranal & Speculoos/Europa & 473   & 12 & 1$\times$1 & I+z & full transit  \\
2018 Apr 15 & La Silla & Eulercam & 193   & 40 & 1$\times$1 & NGTS & full transit \\
2018 Apr 16 & SSO  & \LCO     & 19   & 180 & 1$\times$1 & i  &  egress    \\
2018 Apr 19 & CTIO & \LCO     & 11  & 180 & 1$\times$1 & i  &  ingress    \\
2018 Apr 19 & La Silla & Eulercam & 140   & 55 & 1$\times$1 & NGTS & full transit\\
2018 Apr 20 & SSO  & \LCO     & 16   & 180 & 1$\times$1 & i  &  egress    \\
2018 Apr 22 & SAAO & \SHOC & 2160 & 2 & 4$\times$4 & I & egress\\
2018 Apr 27 & CTIO & \LCO     & 19  & 180 & 1$\times$1 & i  &  egress	     \\
2018 May 06 & SSO  & \LCO     & 15  & 180 & 1$\times$1 & i  &  egress	     \\

		\hline
	\end{tabular}
\end{table*}


\section{Spectroscopy}
\label{sec:spec}
We obtained multi-epoch spectroscopy for \Nstar\ with the HARPS spectrograph \citep{2003Msngr.114...20M} on the ESO 3.6\,m telescope at \LSO, Chile, between 2017 December 01 and 2018 April 10 under programme ID \progid.

We used the standard HARPS data reduction software (DRS) to the measure the radial velocity of \Nstar{} at each epoch. This was done via cross-correlation with the \mask ~binary mask. The exposure times for each spectrum were 2700\,s. The radial velocities are listed, along with their associated error, FWHM, contrast, bisector span and exposure time in Table~\ref{tab:rvs}. 

\begin{figure*}
	\subfloat{\includegraphics[width=1.1\columnwidth,valign=c]{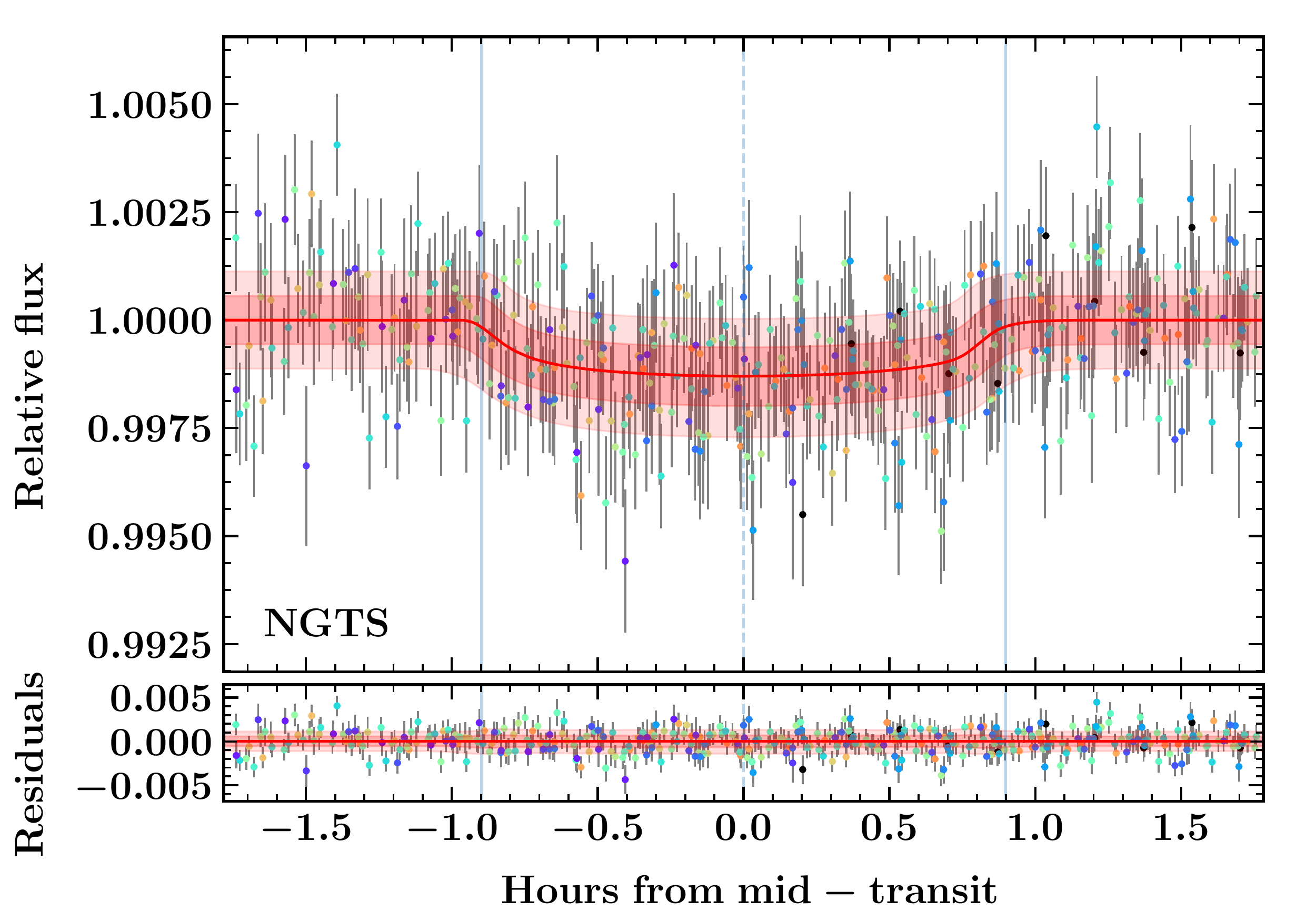}}
    \qquad
    \subfloat{\includegraphics[width=0.9\columnwidth,valign=c]{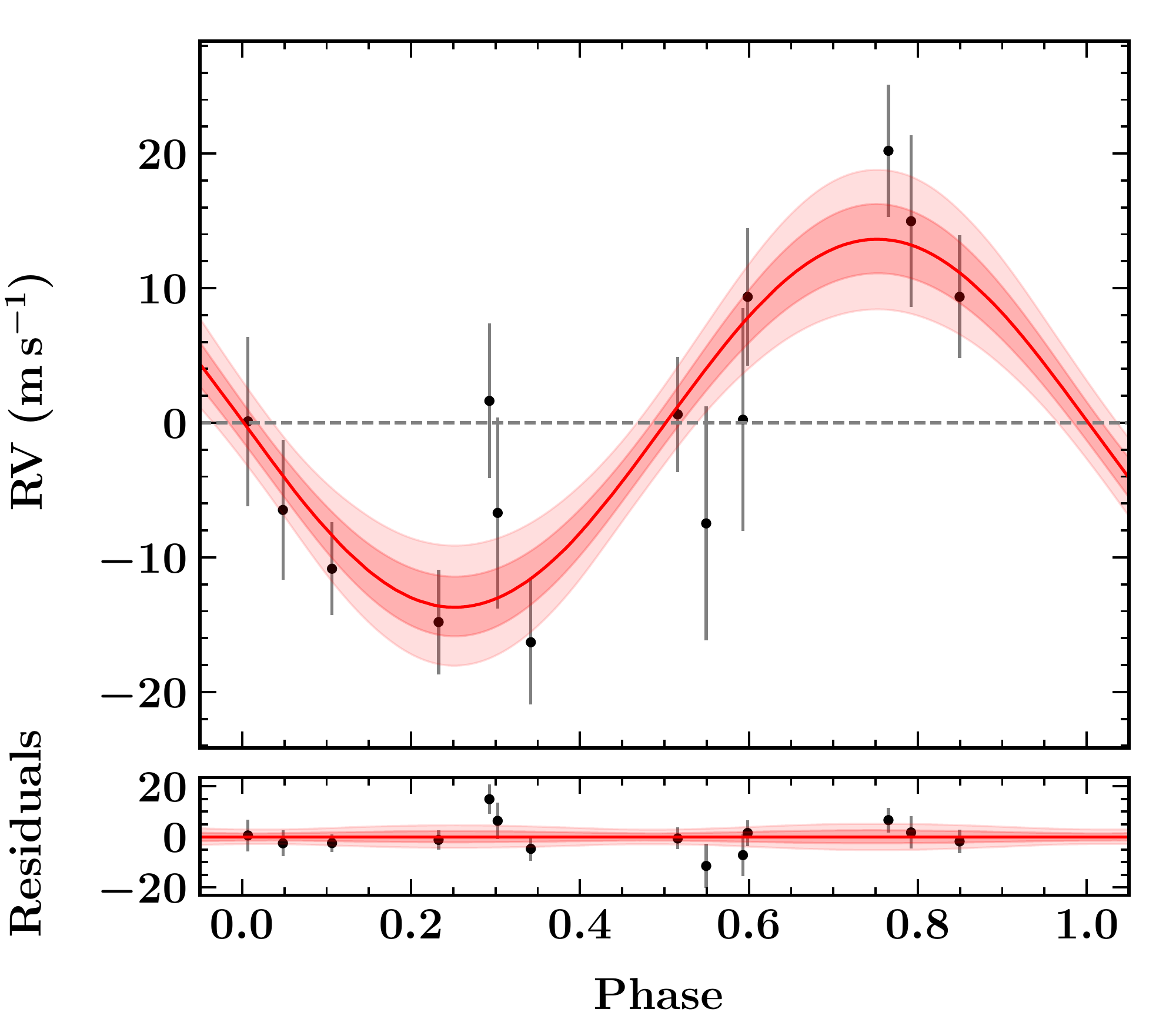}}
    \caption{Left: De-trended NGTS discovery photometry with the \gpe\ model's transit component in red. The data are binned to 10\,min cadence, with individual transits colour-coded. Right: HARPS radial velocity data with the best-fit circular orbit model. In both cases, the red lines and pink shaded regions show the median and the $1\sigma$ and $2\sigma$ confidence intervals of \gpe's posterior model.
}
    \label{fig:global_fit_ngts_rv}
\end{figure*}

\begin{figure}
	\includegraphics[width=\columnwidth]{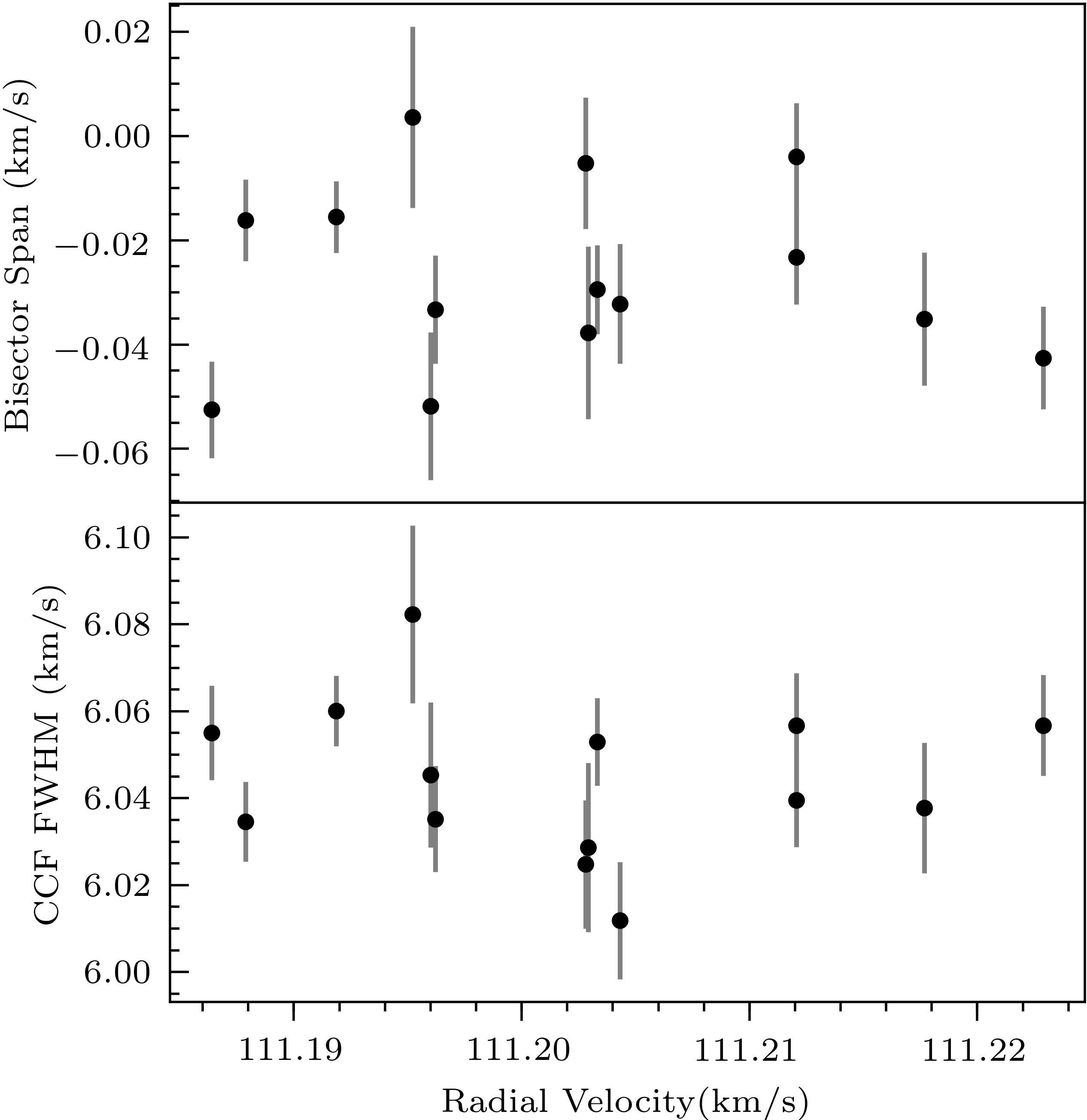}
    \caption{Top: HARPS CCF bisector slopes for the spectra in Table \ref{tab:rvs}, plotted against radial velocity. We find no trend in the bisectors with radial velocity, which can be indicative of a blended system. Bottom: CCF FWHM for the same HARPS spectra. The FWHM of the HARPS CCFs are essentially constant.}
    \label{fig:ccf}
\end{figure}

The \noRV ~radial velocity measurements show a variation in-phase with the period detected by {\sc orion}. With a semi-amplitude of \Nkamp\,\ms~they indicate a Neptune-mass transiting planet (see Figure~\ref{fig:global_fit_ngts_rv}). To ensure that the radial velocity signal originates from a planet orbiting \Nstar{} we analysed the HARPS cross correlation functions (CCF) using the line bisector technique of \citet{queloz2001}. We find no evidence for a correlation between the radial velocity and the bisector spans (see Figure~\ref{fig:ccf}). 

In order to characterise the stellar properties of \Nstar{} we wavelength shift and combine all \noRV\ HARPS spectra to create a high signal-to-noise spectrum for analysis in Section~\ref{sub:stellar}. 

\begin{table*}
	\centering
	\caption{HARPS Radial Velocities for \Nstar{} }
	\label{tab:rvs}
	\begin{tabular}{ccccccc} 
		BJD$_\mathrm{TDB}$			&	RV	& RV err &	FWHM & Contrast & BIS    & Exptime   \\
		(-2,400,000)	& (\kms)& (\kms) &(\kms) &          & (\kms) & (seconds) \\
		\hline
58088.796126 &	111.20332 &	0.00428 &	6.05296 &	42.601 &	-0.02945 & 2700 \\
58090.790319 &	111.20280 &	0.00629 &	6.02478 &	40.836 &	-0.00521 & 2700 \\
58091.804515 &	111.22290 &	0.00492 &	6.05677 &	41.368 &	-0.04256 & 2700 \\
58113.827181 &	111.18789 &	0.00390 &	6.03459 &	42.909 &	-0.01615 & 2700 \\
58125.694569 &	111.19185 &	0.00345 &	6.06011 &	42.884 &	-0.01553 & 2700 \\
58127.682623 &	111.20292 &	0.00828 &	6.02870 &	42.492 &	-0.03770 & 2700 \\
58160.714908 &	111.20432 &	0.00574 &	6.01184 &	42.535 &	-0.03220 & 2700 \\
58162.719953 &	111.21767 &	0.00639 &	6.03778 &	42.665 &	-0.03507 & 2700 \\
58189.543826 &	111.21205 &	0.00456 &	6.03951 &	43.090 &	-0.02321 & 2700 \\
58190.545562 &	111.21205 &	0.00512 &	6.05674 &	42.993 &	-0.00393 & 2700 \\
58199.563729 &	111.18639 &	0.00463 &	6.05506 &	42.778 &	-0.05251 & 2700 \\
58215.559376 &	111.19600 &	0.00710 &	6.04539 &	43.125 &	-0.05181 & 2700 \\
58216.556670 &	111.19622 &	0.00519 &	6.03521 &	43.087 &	-0.03327 & 2700 \\
58218.564257 &	111.19522 &	0.00870 &	6.08229 &	43.049 &	0.00362 & 2700 \\
		\hline
	\end{tabular}
\end{table*}

\section{Analysis}
\label{sec:analysis}


\subsection{Stellar Properties}
\label{sub:stellar}
\subsubsection{Spectroscopic stellar parameters}
The HARPS spectra were analysed using SPECIES \citep{Soto2018}. This is a python tool to derive stellar parameters in an automated way, from high resolution echelle spectra. SPECIES measures the equivalent widths (EWs) for a list of FeI and FeII lines using ARES \citep{Sousa2015}, and they are input into MOOG \citep{sneden_moog}, along with ATLAS9 model atmospheres \citep{ATLAS9}, to solve the radiative transfer equation. The correct set of atmospheric parameters (\teff, \logg, $\left[Fe/H\right]$) are reached when no correlations exist between the obtained abundances for each line, the line excitation potential and the reduced EW (EW/$\lambda$), and the abundance of neutral and ionized iron agree. Mass and radius are found by interpolating through a grid of MIST isochrones \citep{Dotter2016}, using a Bayesian approach. The atmospheric parameters, along with the extinction corrected magnitudes and the Gaia parallax listed in Table~\ref{tab:stellar}, were used as priors. The extinction for each band was computed using the maps from \citet{Bovy2016}. Finally, the rotational and macroturbulent velocity were derived using the relation from \citet{dosSantos2016}, and by line fitting to a set of five absorption lines. SPECIES gives a stellar radius of \NstarradiusSPECIES\,\rsun\ and a metallicity of \NmetalSPECIES ~dex. The stellar parameters measured by SPECIES are listed in Table~\ref{tab:stellar}.
\begin{table}
	\centering
	\caption{Stellar Properties for \Nstar}
	\begin{tabular*}{\columnwidth}{lcc} 
	Property	&	Value		&Source\\
	\hline
    \multicolumn{3}{l}{Astrometric Properties}\\
    R.A.		& \NRA			&2MASS	\\
	Dec			& \NDec			&2MASS	\\
    2MASS I.D.	& \Ntwomass & 2MASS \\
    Gaia source I.D. & \Ngaia & Gaia DR2 \\
    $\mu_{{\rm R.A.}}$ (\masy) & \NpropRA & Gaia DR2 \\
	$\mu_{{\rm Dec.}}$ (\masy) & \NpropDec & Gaia DR2 \\
    $\gamma_{{\rm gaia}}$ (\kms) & \Ngaiagamma & Gaia DR2 \\
    Parallax (mas)	& \Nparallax		&Gaia DR2	\\    
    Distance (parsec) & \Ndist & Gaia DR2 \\
    \\
    \multicolumn{3}{l}{Photometric Properties}\\
	V (mag)		&\NVmag 	&APASS\\
	B (mag)		&\NBmag		&APASS\\
	g (mag)		&\Ngmag		&APASS\\
	r (mag)		&\Nrmag		&APASS\\
	i (mag)		&\Nimag		&APASS\\
    G (mag)		&\NGAIAmag	&Gaia DR2\\
    $\text{G}_{\text{RP}}$ (mag) &\NGAIARPmag	&Gaia DR2\\
    $\text{G}_{\text{BP}}$ (mag) &\NGAIABPmag	&Gaia DR2\\
    NGTS (mag)	&\NNmag		&this work\\
    J (mag)		&\NJmag		&2MASS	\\
   	H (mag)		&\NHmag		&2MASS	\\
	K (mag)		&\NKmag		&2MASS	\\
    W1 (mag)	&\NWmag		&WISE	\\
    W2 (mag)	&\NWWmag	&WISE	\\
    W3 (mag)	&\NWWWmag	&WISE	\\
    \\
    \multicolumn{3}{l}{Derived Properties}\\
    T$_{\rm eff}$ (K)    & \NteffSPECIES        & SPECIES \\
    $\left[M/H\right]$   & \NmetalSPECIES       & SPECIES\\
    $v\sin i$ (\kms)	     & \NvsiniSPECIES       & SPECIES\\
    $\gamma_{RV}$ (\kms) & \Ngamma       & Global modelling \\
    $\log g$                & \NloggSPECIES        & SPECIES \\
    \mstar (\msun)       & \NstarmassSPECIES    & SPECIES \\
    \rstar (\rsun)       & \NstarradiusSPECIES  & SPECIES \\
    $\rho_*$ (\gccc)       & \Nstardensity & SPECIES \\
    \vmac				 & \NvmacSPECIES & SPECIES \\
    Distance (pc)        & \Ndist        & Gaia DR2 \\
	\hline
    \multicolumn{3}{l}{2MASS \citep{2MASS}; WISE \citep{WISE};}\\
    \multicolumn{3}{l}{APASS \citep{APASS};}\\
    \multicolumn{3}{l}{Gaia DR2 \citep{gaia_dr2}}\\
    \multicolumn{3}{l}{ and \citet{Mann15}}
	\end{tabular*}
    \label{tab:stellar}
\end{table}

\subsubsection{Kinematics and Environment}
\label{sub:kinematics}

Using the Gaia parallax and the tables of \citet{Bailer-Jones18} we estimate the distance to \Nstar\ to be \Ndist\,parsec.

\Nstar\ has a relatively high proper motion, consistent with what we expect for the estimated distance and spectral type, of \NpropRA\,\masy\ and \NpropDec\,\masy\ in R.A. and Dec. respectively.  It has a very high systemic radial velocity as determined from our HARPS observations (\Ngamma\,\kms) and confirmed from Gaia DR2 (\Ngaiagamma\,\kms).

When combined with the Gaia DR2 parallax \citep{gaia_dr2}, we derived the following Galactic velocity components ($U_{LSR}, V_{LSR}, W_{LSR}$) with respect to the Local Standard of Rest to be ($66.99 \pm 0.10, -72.46 \pm 0.15, -38.22 \pm 0.08$)\,\kms, assuming the Local Standard of Rest is UVW=(11.1, 12.24, 7.25)\,\kms\ from \cite{lsr-2010}. This suggests that \Nstar\ is a member of the thick disk population (V$_{\rm tot}=105.8$ \kms, see e.g. \citealp{gaia_dr2_hr}) similar to NGTS-1 \citep{bayliss2018}.

There are no other sources within 15\arcsec\ of \Nstar\ in the Gaia DR2 catalogue.  This means we can rule out any blended object down to a Gaia magnitude of approximately G=20.7 beyond 2\arcsec\ and within the NGTS photometric aperture. However, the Gaia DR2 completeness for close companions falls off within 2\arcsec\ and is zero within 0.5\arcsec\ \citep{arenou}.


\subsection{Global Modelling}
\label{sub:global}

To obtain fundamental parameters for \Nplanet, we modelled the light and radial velocity curves of \Nstar\ using \gpe\ \citep{Gillen17}. These comprise the NGTS discovery light curve, 12 follow-up light curves from six 1m-class telescopes, and 14 HARPS RVs (as set out in Sect.~\ref{sec:ngts}, \ref{sec:phfu} and \ref{sec:spec}). \gpe\ comprises a central transiting planet and eclipsing binary model, which is coupled with a Gaussian process (GP) model, and wrapped within an MCMC. Limb darkening (LD) is incorporated using the analytic method of \citet{Mandel02} for the quadratic law with the profiles and uncertainties constrained by the predictions of LDtk \citep{Parviainen15}. 

Most stars are intrinsically variable, which can affect the apparent shape and depth of planet transits. The more active the star, or the higher the level of instrumental systematics, or the shallower the transit signal, the greater the effect on the transit modelling and hence the inferred planet parameters. \Nstar\ is a relatively quiet star and the NGTS systematics are low, but the transit signal is very shallow. Furthermore, the 12 follow-up light curves obtained from 6 facilities all have their own level of systematics and hence correlated noise. \gpe\ is designed to propagate the effect of variability/systematics into the inferred stellar and planet properties. The reader is referred to \citet{Gillen17} for further details on the model. 

We modelled the orbit of \Nplanet\ assuming both a circular and an eccentric orbit about the host star. Both models were identical except for the orbital eccentricity constraints. Each light curve was given its own variability/systematics model (with the exception of the LCO light curves which, given the observational uncertainties, all shared the same GP variability model). A Matern-32 kernel was chosen for all light curves given the low level of apparent stellar variability but clear presence of instrument systematics and/or atmospheric variability. Limb darkening profiles were generated using LDtk, given estimates of \teff, \logg\ and \FeH\ from SPECIES (see Sect.~\ref{sub:stellar}). The LD uncertainty was inflated by a factor of 10 to account for systematic uncertainties in stellar atmosphere models around where \Nstar\ lies. The NGTS light curve was binned to 10\,min cadence and all other light curves to 3\,min cadence, with the \gpe\ model integrated accordingly. The 14 HARPS RVs were modelled with a Keplerian orbit where the uncertainties were allowed to inflate, if required. We ran the MCMC for 80\,000 steps with 200 walkers, discarding the first 30\,000 points as burn in and using a thinning factor of 500. 

We find that the derived planet parameters from both the circular and eccentric models are consistent to within their $1\sigma$ uncertainties. Furthermore, the eccentric model converges on an eccentricity consistent with zero at the $1.5\sigma$ level, which suggests that there is no clear evidence for an eccentric orbit in our data. We therefore adopt the circular model as our main model. 

\begin{figure*}
\centering
	\subfloat{\includegraphics[width=0.33\textwidth]{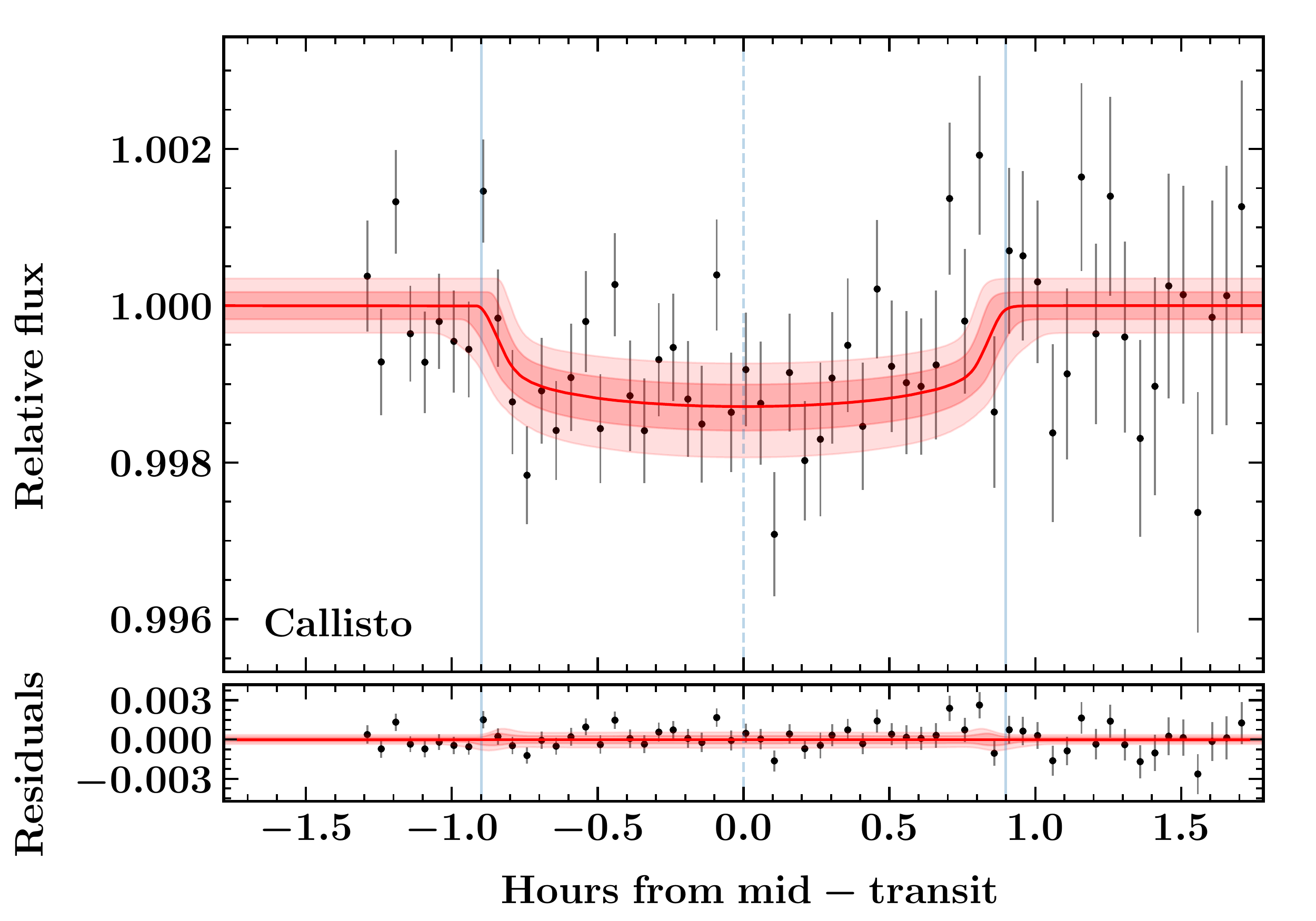}}
\subfloat{\includegraphics[width=0.33\textwidth]{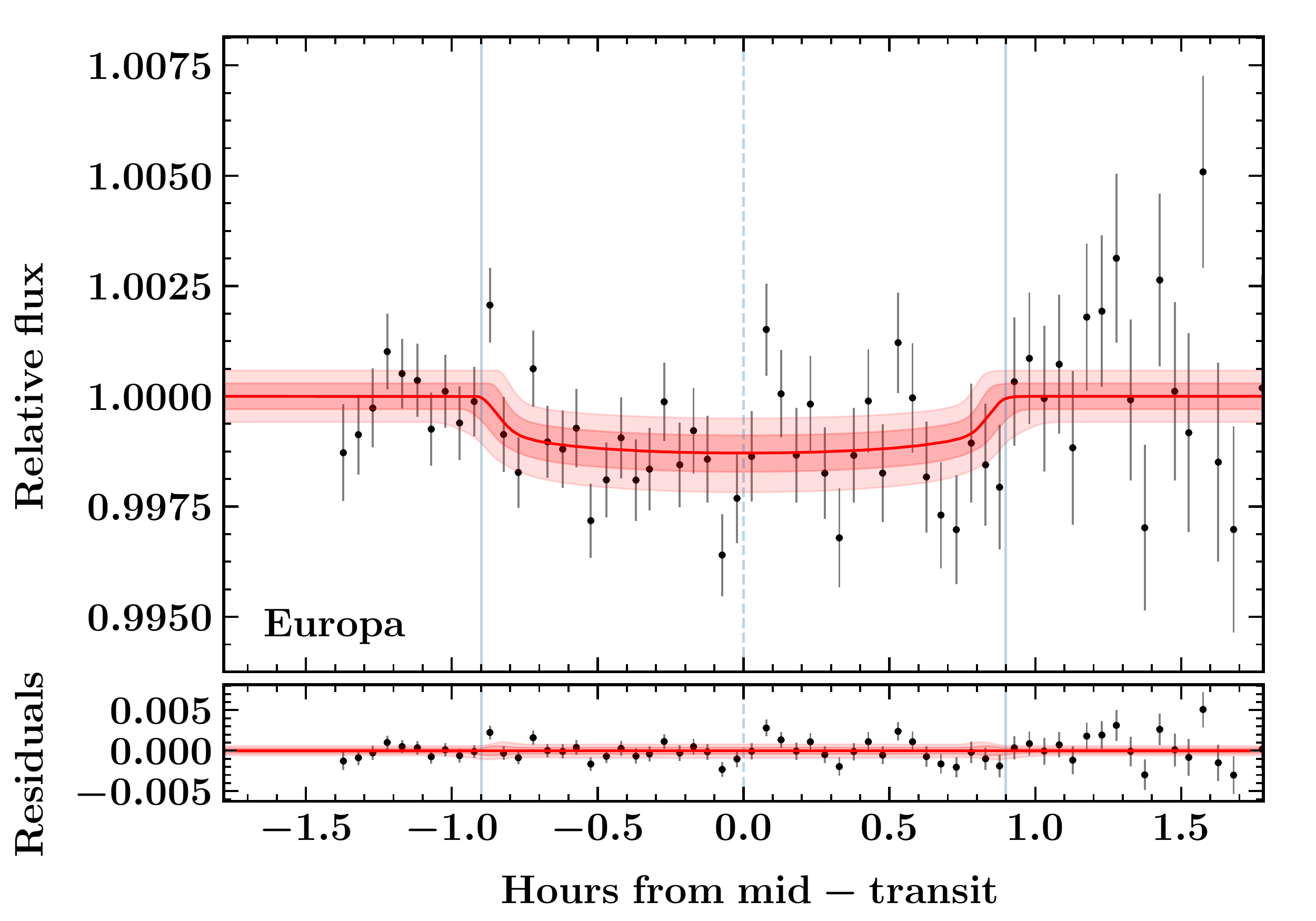}}
\subfloat{\includegraphics[width=0.33\textwidth]{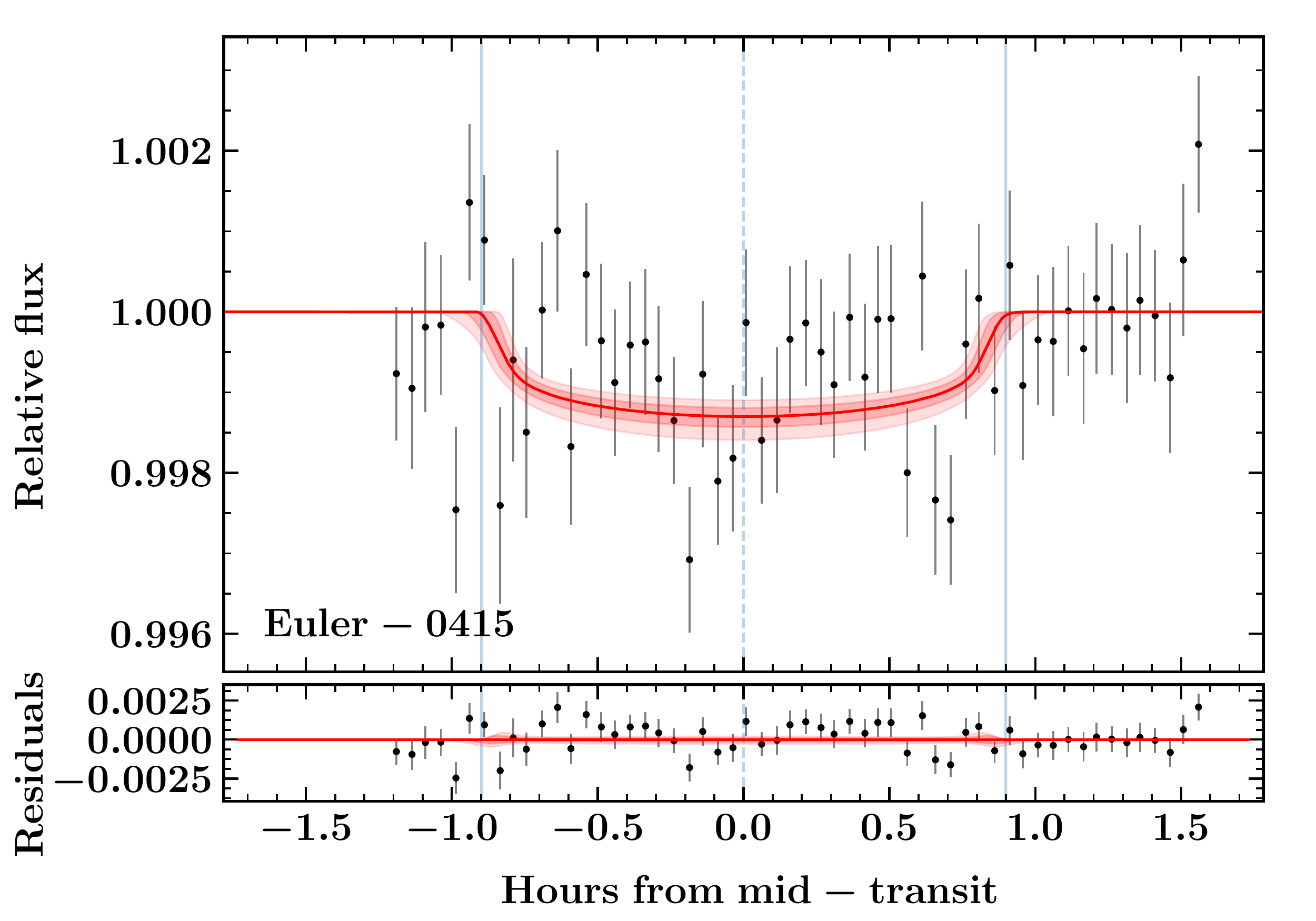}}
\\
\subfloat{\includegraphics[width=0.33\textwidth]{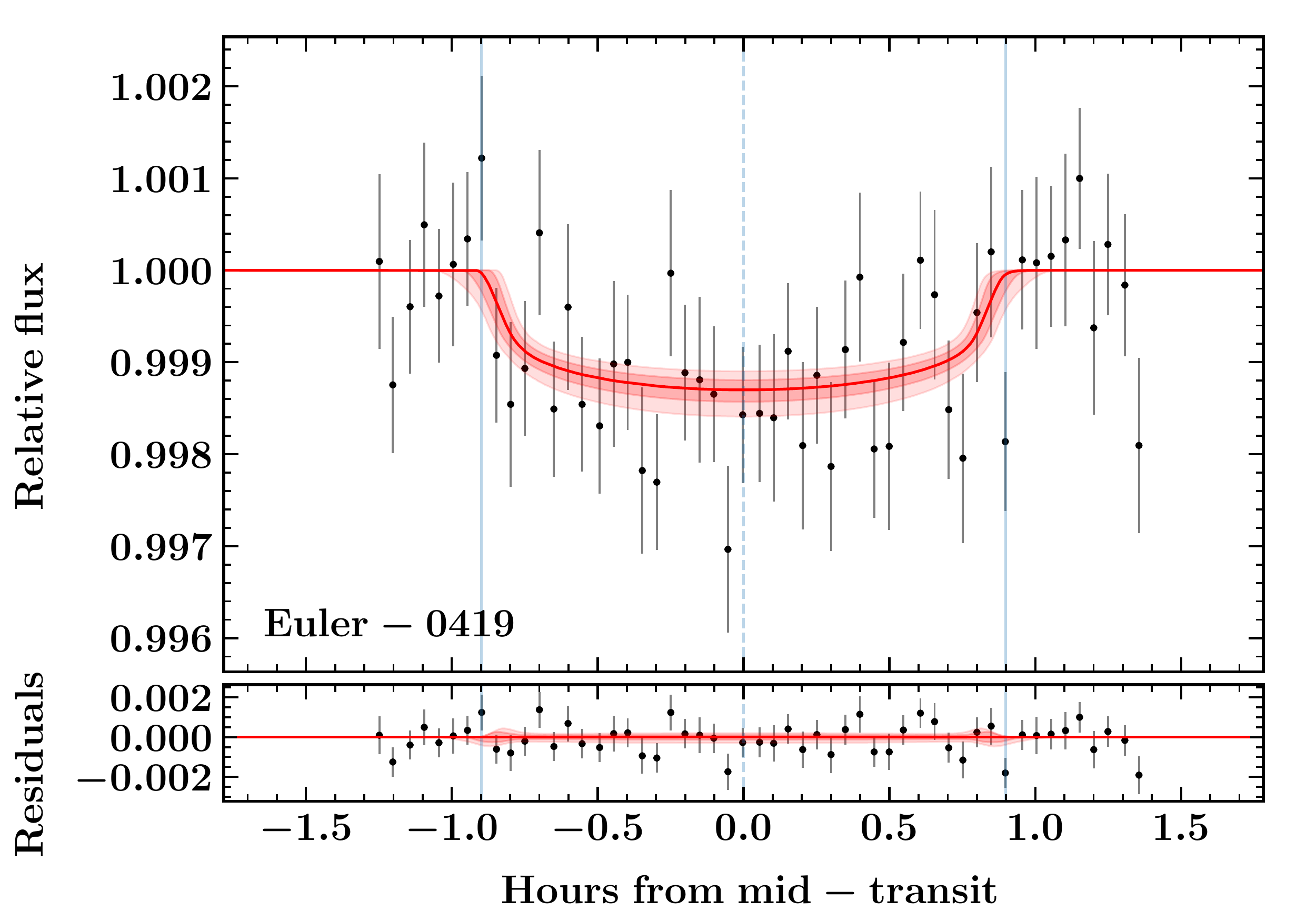}}
\subfloat{\includegraphics[width=0.33\textwidth]{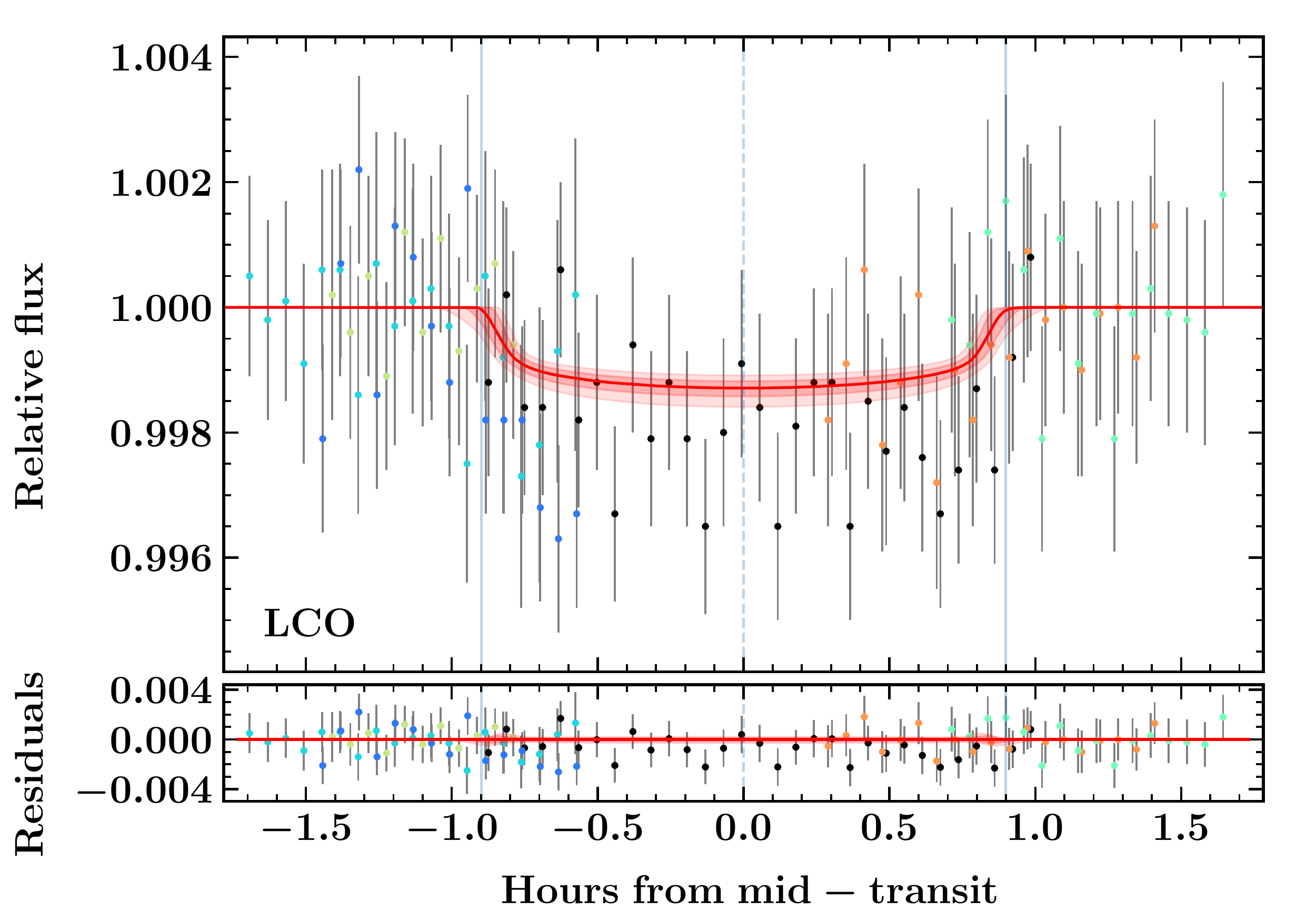}}
\subfloat{\includegraphics[width=0.33\textwidth]{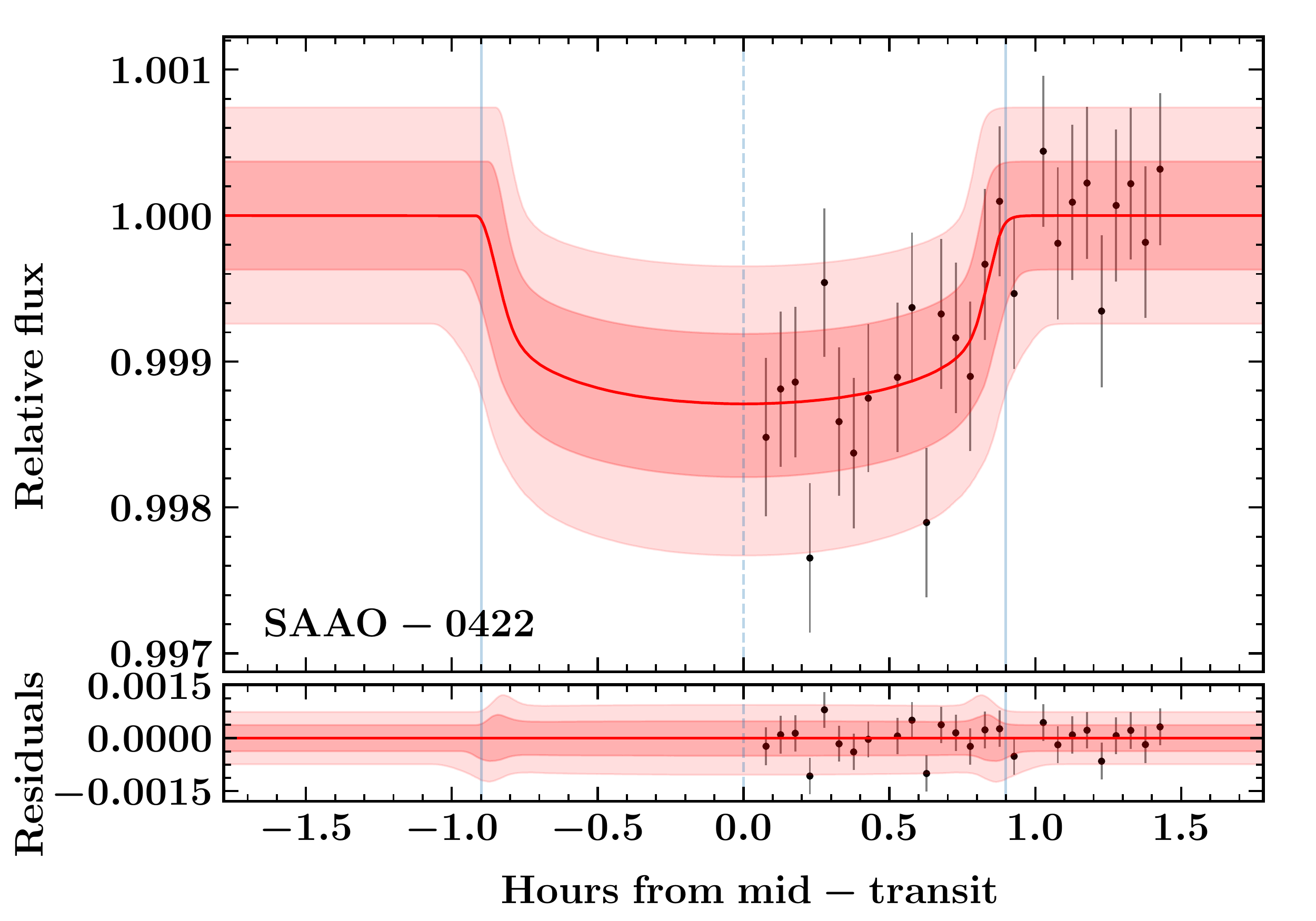}}
\\
\subfloat{\includegraphics[width=0.33\textwidth]{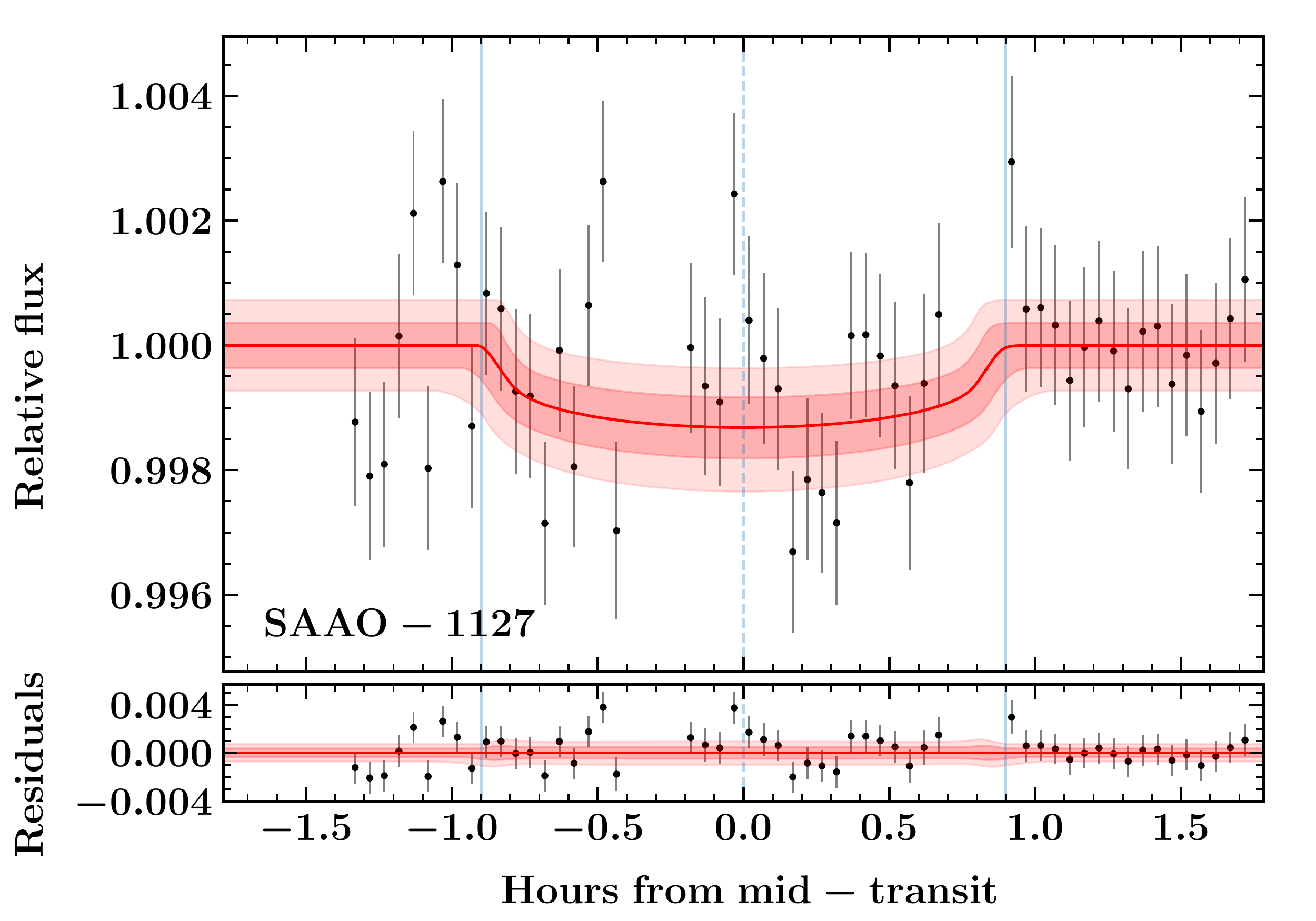}}

\caption{De-trended photometry from the SPECULOOS, EULER, LCO and SAAO follow-up observations, plotted with the \gpe\ model's transit component. The red lines and pink shaded regions show the median and the $1\sigma$ and $2\sigma$ confidence intervals of \gpe's posterior model.}
\label{fig:global_fit_fup_phot}
\end{figure*}

We find that \Nplanet\ comprises a \Nmass\,\mearth\ and \Nradius\,\rearth\ planet, with a corresponding density of \Ndensity\,\gccc , which orbits \Nstar\ in \Nperiod\,d with a semi-major axis of \Nau\,AU. Fitted and derived parameters of the \gpe\ model are reported in Table \ref{tab:planet}. 
The best-fit \gpe\ models are plotted against the de-trended NGTS discovery photometry and the HARPS radial velocity data are presented in Figure~\ref{fig:global_fit_ngts_rv}, and against the de-trended 1\,m-class follow-up photometry in Figure~\ref{fig:global_fit_fup_phot}. The light curve data in these plots has been de-trended with respect to \gpe's variability model and, accordingly, the \gpe\ model displayed is the posterior transit component alone.

\begin{table}
	\centering
	\caption{Planetary properties for \Nplanet{} for a circular orbit and eccentric orbit. We adopt the circular model as the most likely solution, the parameters from the eccentric model are provided for information only.}
	\begin{tabular}{lcc} 
	Property	&	Value (circular) &	Value (ecc)\\
	\hline
    P (days)		&	\Nperiod & $1.3373506\pm0.000008$	\\
	T$_C$ (HJD)		&	\Ntc & $2457607.9978\pm0.0033$	\\
    T$_{14}$ (hours) & \Nduration & $1.79\pm0.09$\\
    $a/{\mathrm R}_{*}$		& \Naoverr & $4.22\pm1.18$\\
	K (\ms) 	&\Nkamp	& $14.0\pm2.0$\\
    e 			& 0.0 (fixed) & $0.14^{+0.18}_{-0.10}$ \\
    $\omega$ (deg) & 0.0 (fixed) & $69^{+35}_{-92}$ \\
    \mpl (\mearth)& \Nmass	& $20.8\pm3.4$\\
    \rpl (\rearth)& \Nradius & $3.18\pm0.27$ \\
    $\mathrm{R}_p/\mathrm{R}_*$ & $0.035\pm0.003$ & $0.035\pm0.003$ \\
    $\rho_{p}$ (\gccc) & \Ndensity & $3.50\pm0.95$\\
    $\rho_{*}$ (\gccc) & $1.91\pm0.16$ & $1.91\pm0.16$\\
    a (AU) & \Nau & $0.017\pm0.004$ \\
    \teq (K) & \NTeq & \NTeq	\\
    $i$ (deg) & \Ninclination & $81.0\pm7.7$ \\
	\hline
	\end{tabular}
    \label{tab:planet}
\end{table}

In addition to the circular fit, we also present the results of the eccentric model fit in Table~\ref{tab:planet}. We suspect that the fitted non-zero eccentricity in this model is due to the sparsity of RV coverage at an orbital phase of $\sim$0.9. Nevertheless, given that the orbit of such a short-period planet as \Nplanet\ would be expected to have circularised, an eccentric orbit if true would be potentially interesting.

\section{Discussion}
\label{sec:discussion}
\begin{figure}
	\includegraphics[width=8.5cm]{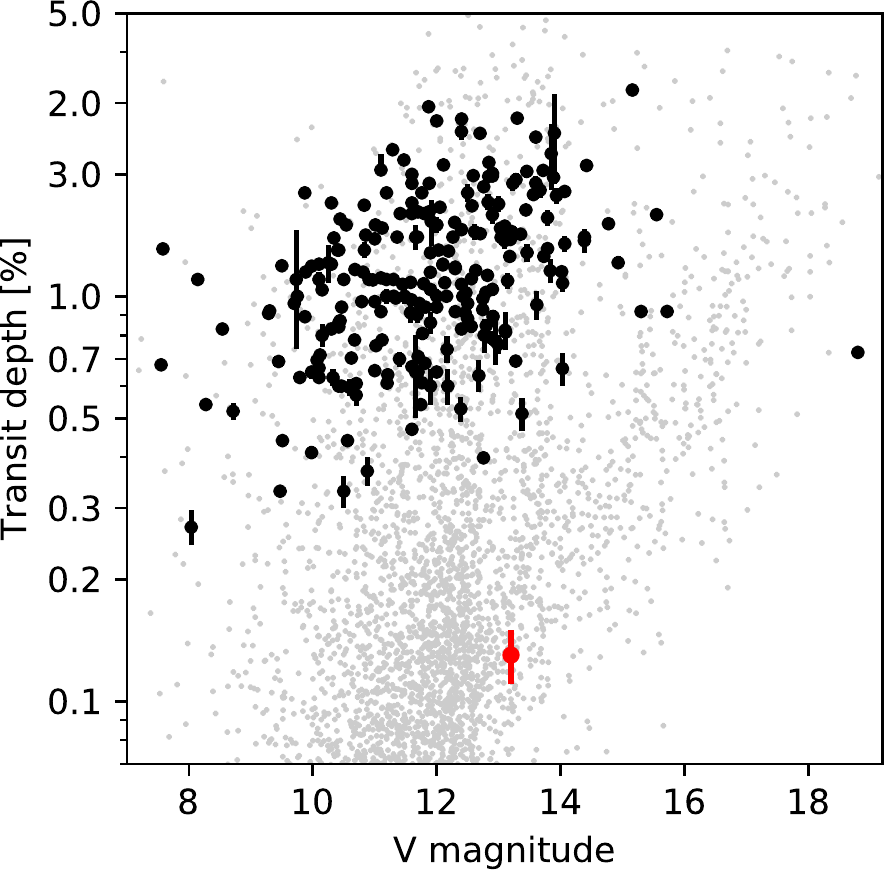}
    \caption{Transit depth versus host star brightness for all transiting exoplanets discovered by wide-field ground-based transit surveys. \Nplanet\ is marked in red. Data from NASA Exoplanet Archive \citep{akeson13} accessed on 2018 May 10. The grey dots show the simulated distribution of planet detections from \TESS\  \citep{barclay2018}.}
    \label{fig:depth}
\end{figure}

\Nplanet\ is the shallowest transiting exoplanet so far discovered from the ground (see Figure~\ref{fig:depth}), with a transit depth of just \Ndepth\,\%.  It is approximately 30\% shallower than the second shallowest discovery - KELT-11b \citep{gaudi17}. The ability to be able to detect such shallow transits allows \NGTS\ to reach down into the Neptunian desert, as evidenced by \Nplanet, in a way that has not previously been possible for ground-based surveys.  It is also encouraging for prospects of following up shallow \TESS\ discoveries using the \NGTS\ facility.

\begin{figure}
\centering
\includegraphics[width=8.5cm]{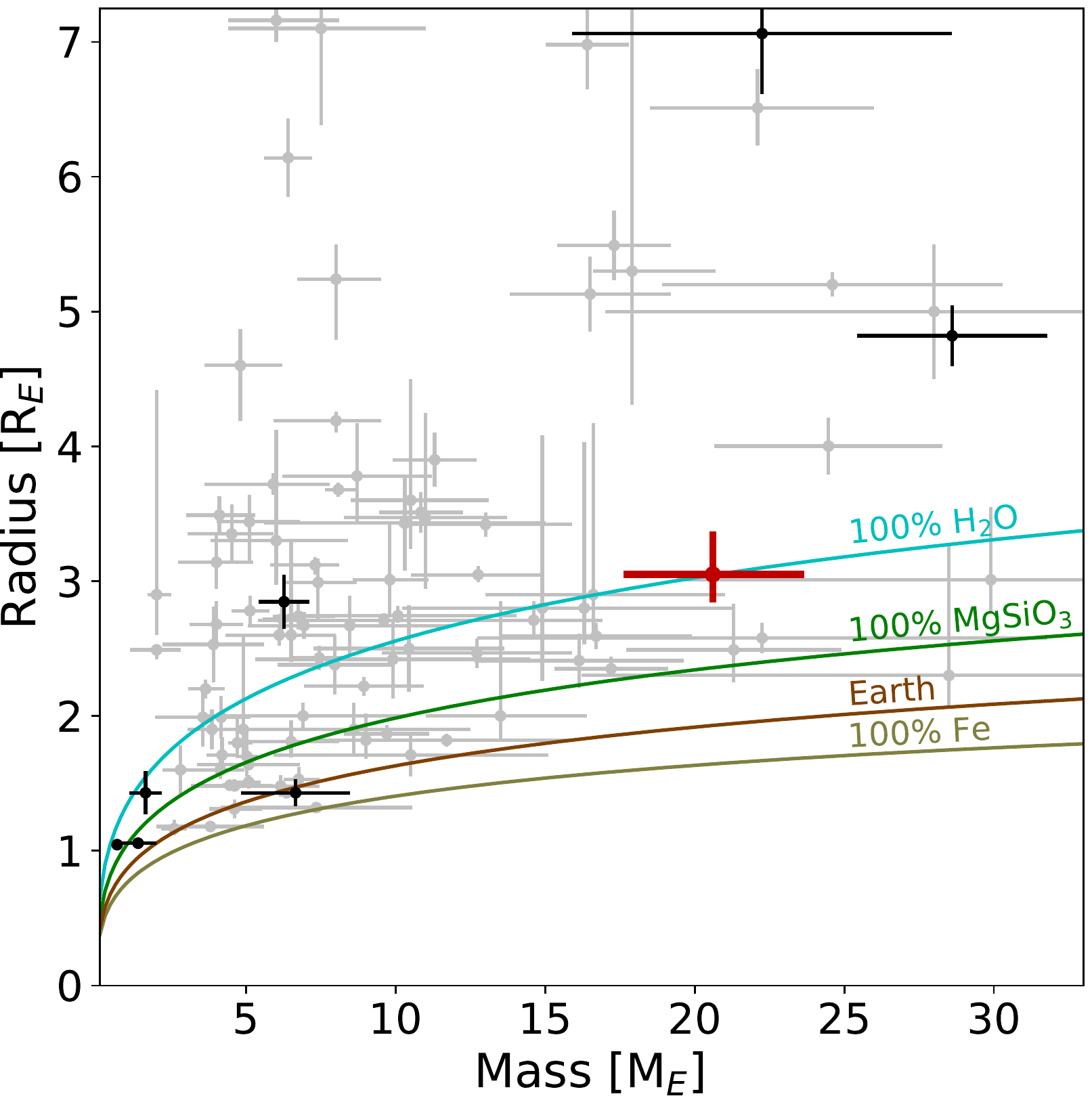}
\caption{The mass and radius for all known transiting planets that have fractional errors on the measured planet mass better than $30$\,\%. The black and grey points show discoveries from ground-based and space-based telescopes respectively. The coloured lines show the theoretical mass-radius relation for solid exoplanets of various compositions \citep{seager2007}. NGTS-4b is highlighted in red.}
\label{fig:mass_radius}
\end{figure}

Figure~\ref{fig:mass_radius} shows the masses and radii of known transiting planets that have masses measured to better than 30\%, along with mass-radius relations from the models of \citet{seager2007}. The mass and radius of \Nplanet\ as measured in this work are consistent with a composition of 100\% H$_2$O, however this is likely to be unphysical given the proximity to the host star, and it is more likely to consist of a rocky core with a water and/or gaseous envelope.

\begin{figure}
	\includegraphics[width=\columnwidth]{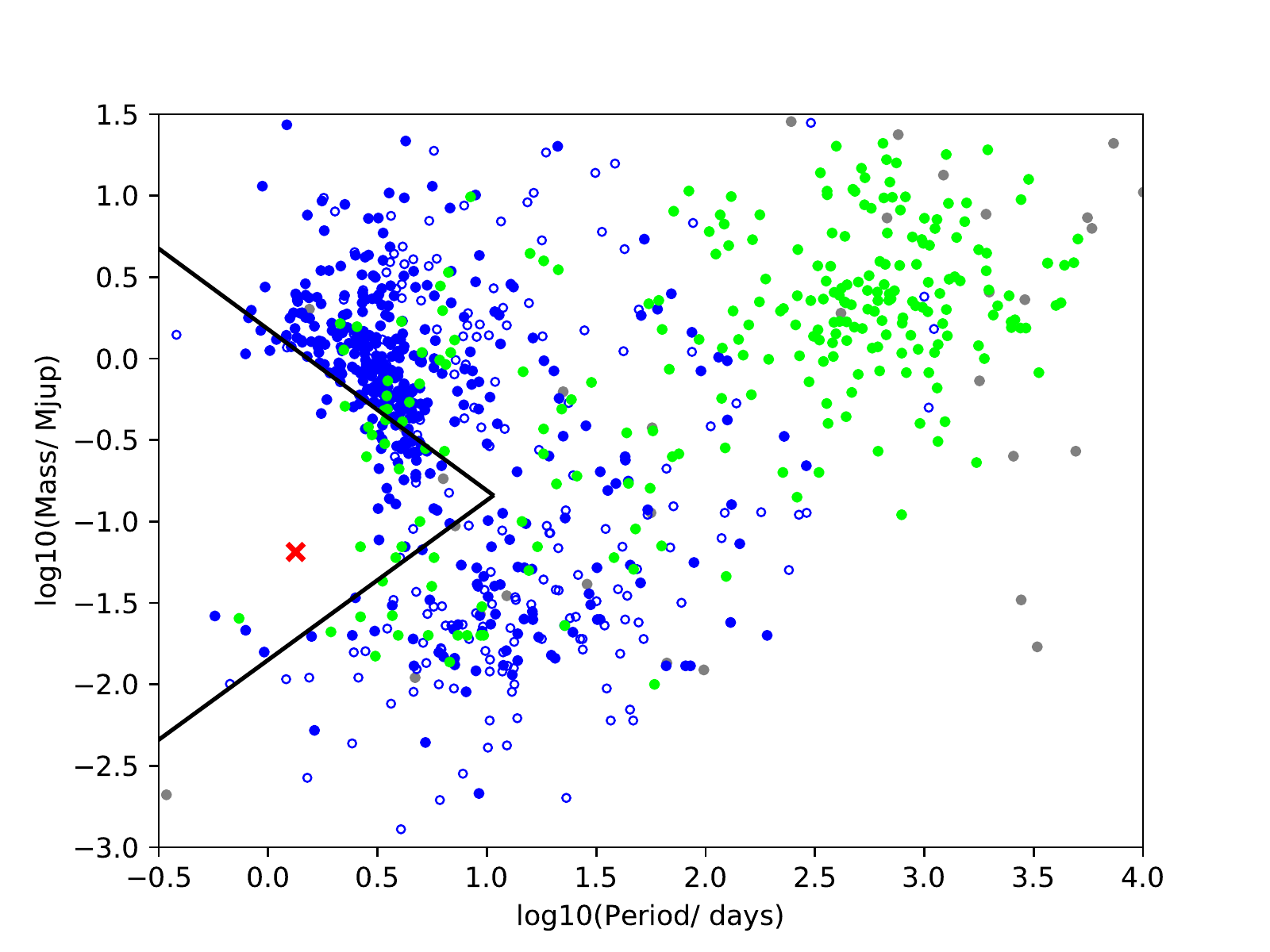}
    \caption{Distribution of mass versus orbital period for planets with a measured mass. Planets discovered by the transit method are shown in blue, with those having masses measured to better than 30\% represented as filled circles. Planets found by radial velocity are shown as green circles and those detected by other methods are shown by grey circles. Black lines represent the Neptunian desert as defined in \citet{mazeh2016}. \Nplanet\ is shown as a red cross. Taken from the NASA Exoplanet Archive on 2018 August 14. 
    }
    \label{fig:periodmass}
\end{figure}


Studies have reported a significant dearth of Neptune-sized ($R<5$\,\rearth ) planets in close orbits (P$<3$\,d), the so-called ``Neptunian desert" \citep{mazeh2016}, perhaps due to the X-ray/EUV flux from the host stars quickly stripping these planets of their atmospheres and leaving them as lower-mass rocky cores. However as can been seen from Figure \ref{fig:periodmass}, \Nplanet\ is clearly in a central region of the Neptunian desert, and is likely to still contain a significant atmosphere despite its proximity to its host star.  There is nothing in our photometric or spectroscopic data to suggest that \Nstar\ is particularly young, so it is unlikely that this can explain the existence of \Nplanet\ in the Neptunian desert. 

Following the method of \citet{King18}, and assuming a canonical evaporation efficiency of 15 per cent and the X-ray-age relations of \citet{Jackson12}, we estimate a mass loss rate of 10$^{10}$\,g\,s$^{-1}$, even at an assumed age of 5\,Gyr. This is at least an order of magnitude higher than the inferred mass loss rate of the Neptune GJ 436b, which was observed to have a 56\% deep transit in Lyman-alpha,  corresponding to an extended comet-like tail of evaporating material \citep{Ehrenreich15,Lavie17}. 

The X-ray luminosity of the star will have been two orders of magnitude higher during its early evolution, when it was maximally active \citep[e.g.][]{Jackson12}. NGTS-4b may have survived in the Neptunian desert due to an unusually high core mass \citep[e.g.][]{Owen2018}, or it might have migrated to its current close-in orbit after this epoch of maximum stellar activity \citep[e.g.][]{Jackson12}.

Future discoveries from \NGTS\ and \TESS\ of more Neptune-sized exoplanets should allow us to more carefully characterise the Neptunian desert and the systems that reside within it. The \TESS\ mission \citep{ricker2014} is set to deliver a large number of transiting exoplanets, the bulk of which will be much shallower than can be detected from ground-based surveys (see Figure~\ref{fig:depth}).  However the discovery of \Nplanet\ shows that the \NGTS\ facility is able to detect shallow transits in the magnitude range where many of the \TESS\ candidates reside.  This will be particularly important for follow-up of single-transit candidates.  \citet{villanueva18} estimate over 1000 single-transit candidates from \TESS\, of which 90\% will be deeper than 0.1\%. Such candidates will be amenable to follow-up with \NGTS.  

\section{Conclusions}
\label{sec:conclusions}
We have presented the discovery of NGTS-4b, a sub-Neptune-sized transiting exoplanet located within the Neptunian Desert. The discovery of \Nplanet\ is a breakthrough for ground-based photometry; the \Ndepth\ per cent transit being the shallowest ever detected from a wide-field ground-based photometric survey. It allows us to begin to probe the Neptunian desert and find rare exoplanets that reside in this region of parameter space. In the near future, such key systems will allow us to place constraints on planet formation and evolution models and allow us to better understand the observed distriubution of planets. Together with future planet detections by NGTS and \TESS\ we will get a much clearer view on where the borders of the Neptunian desert are and how they depend on stellar parameters.

\section*{Acknowledgements}
Based on data collected under the NGTS project at the ESO La Silla Paranal Observatory.  The NGTS facility is operated by the consortium institutes with support from the UK Science and Technology Facilities Council (STFC)  project ST/M001962/1.
This paper uses observations made at the South African Astronomical Observatory (SAAO). 
The contributions at the University of Warwick by PJW, RGW, DLP, DJA, BTG and TL have been supported by STFC through consolidated grants ST/L000733/1 and ST/P000495/1. 
Contributions at the University of Geneva by DB, FB, BC, LM, and SU were carried out within the framework of the National Centre for Competence in Research ``PlanetS" supported by the Swiss National Science Foundation (SNSF).
The contributions at the University of Leicester by MRG and MRB have been supported by STFC through consolidated grant ST/N000757/1.
CAW acknowledges support from the STFC grant ST/P000312/1.
EG gratefully acknowledges support from Winton Philanthropies in the form of a Winton Exoplanet Fellowship. 
JSJ acknowledges support by Fondecyt grant 1161218 and partial support by CATA-Basal (PB06, CONICYT).
DJA gratefully acknowledges support from the STFC via an Ernest Rutherford Fellowship (ST/R00384X/1).
PE and HR acknowledge the support of the DFG priority program SPP 1992 "Exploring the Diversity of Extrasolar Planets" (RA 714/13-1).
LD acknowledges support from the Gruber Foundation Fellowship.
The research leading to these results has received funding from the European Research Council under the FP/2007-2013 ERC Grant Agreement number 336480 and from the ARC grant for Concerted Research Actions, financed by the Wallonia-Brussels Federation. This work was also partially supported by a grant from the Simons Foundation (PI Queloz, ID 327127).
This work makes use of observations from the LCOGT network.
This work has made use of data from the European Space Agency (ESA)
mission {\it Gaia} (\url{https://www.cosmos.esa.int/gaia}), processed by
the {\it Gaia} Data Processing and Analysis Consortium (DPAC,
\url{https://www.cosmos.esa.int/web/gaia/dpac/consortium}). Funding
for the DPAC has been provided by national institutions, in particular
the institutions participating in the {\it Gaia} Multilateral Agreement.
PyRAF is a product of the Space Telescope Science Institute, which is operated by AURA for NASA.
This research has made use of the NASA Exoplanet Archive, which is operated by the California Institute of Technology, under contract with the National Aeronautics and Space Administration under the Exoplanet Exploration Program.



\bibliographystyle{mnras}
\bibliography{paper} 








\bsp	
\label{lastpage}
\end{document}